\documentclass[aip,jcp,12pt,reprint,floatfix]{revtex4-1}

\pdfoutput=1

\usepackage{graphicx}
\usepackage[version=3]{mhchem}

\begin{document}

\title{Photoelectron Spectra of 2-Thiouracil, 4-Thiouracil and 2,4-Dithiouracil}

\author{Matthias Ruckenbauer}
\affiliation{Institute of Theoretical Chemistry, University of Vienna, W\"ahringer Stra\ss{}e 17, 1090 Vienna, Austria}

\author{Sebastian Mai}
\affiliation{Institute of Theoretical Chemistry, University of Vienna, W\"ahringer Stra\ss{}e 17, 1090 Vienna, Austria}

\author{Philipp Marquetand}
\email{philipp.marquetand@univie.ac.at}
\affiliation{Institute of Theoretical Chemistry, University of Vienna, W\"ahringer Stra\ss{}e 17, 1090 Vienna, Austria}

\author{Leticia Gonz\'alez}
\email{leticia.gonzalez@univie.ac.at}
\affiliation{Institute of Theoretical Chemistry, University of Vienna, W\"ahringer Stra\ss{}e 17, 1090 Vienna, Austria}

%%=================================================================================================
%%=================================================================================================

\date{\today}
\begin{abstract}
Ground- and excited-state UV photoelectron spectra of thiouracils (2-thiouracil, 4-thiouracil and 2,4-dithiouracil) have been simulated using multireference configuration interaction calculations and Dyson norms as measure for the photoionization intensity. 
Except for a constant shift, the calculated spectrum of 2-thiouracil agrees very well with experiment, while no experimental spectra are available for the two other compounds. 
For all three molecules, the photoelectron spectra show distinct bands due to ionization of the sulphur and oxygen lone pairs and the pyrimidine $\pi$ system.
The excited-state photoelectron spectra of 2-thiouracil show bands at much lower energies than in the ground state spectrum, allowing to monitor the excited-state population in time-resolved UV photoelectron spectroscopy (TRPES) experiments.
However, the results also reveal that single-photon ionization probe schemes alone will not allow monitoring all photodynamic processes existing in 2-thiouracil. 
Especially, due to overlapping bands of singlet and triplet states the clear observation of intersystem crossing will be hampered.
\end{abstract}

%%=================================================================================================
%%=================================================================================================

% \pacs{32.10.Hq., 32.80.-t, 82.50.-m, 82.50.Hp, 82.53.-k}
%%% 82.50.-m      Photochemistry
%%% 82.53.-k      Femtochemistry
%%% 82.50.Hp      Processes induced by visible and UV light
%%% 32.10.Hq   Ionization potentials, electron affinities
%%% 32.80.-t   Photoionization and excitation
\maketitle

%%=================================================================================================
%%=================================================================================================

\section{Introduction}\label{sec:intro}

Thionucleobases, which are nucleobase analogues where one or more oxygen atoms have been substituted by sulphur, play an important role in molecular biology and pharmacology. 
They are, e.g., contained in natural t-RNA\cite{Zachau1969,Lipsett1965} and, given their structural similarity to canonical nucleobases, they can be used as fluorescent probes in DNA and RNA,\cite{Favre1998JPPB,Favre1990} as well as for pharmaceutical purposes such as photosensitizers for photodynamic therapy\cite{Reelfs2012,Massey2001} or antithyroid drugs\cite{Cooper2005NEJM}.
Besides their occurrence in biological environments and wide applicability, thiobases have received a lot of attention due to their remarkably different photophysics in comparison to canonical nucleobases. 
Despite their similar structure, the  presence of a thiocarbonyl group results in a red-shift of the UV absorption spectrum and an increase of intersystem crossing.\cite{Igarashi-Yamamoto1981BBA,Pollum2014TCC,Pollum2014JCP} 
These differences have prompted a large number of studies with focus on their electronic structure\cite{Cui2013JCP,Gobbo2014CTC,Mai2015JPCA,Martinez-Fernandez2012CC} and excited-state dynamics, both from the theoretical\cite{Martinez-Fernandez2014CS} and experimental 
points of view, in particular by means of transient absorption spectroscopy.\cite{Vendrell-Criado2013PPS,Pollum2014JACS,Pollum2014JCP,Pollum2014TCC,Taras-Goslinska2014JPPA,Pollum2015PCCP}

\begin{figure}
\includegraphics[scale=1]{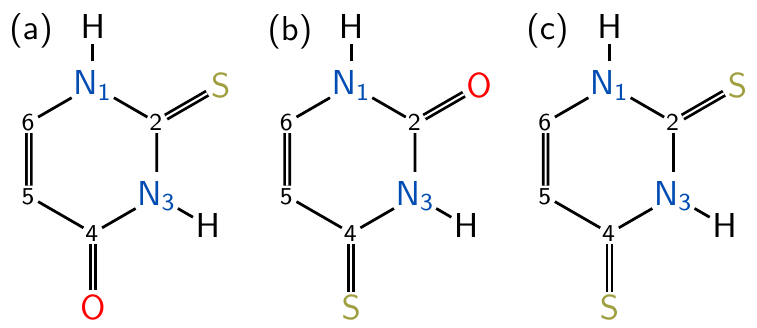}
     \caption{Structures of (a) 2-thiouracil (2TU), (b) 4-thiouracil (4TU) and (c) 2,4-dithiouracil (DTU), with numbering of the ring atoms. 
     Hydrogen atoms attached to carbons are omitted.}
     \label{fig:structures}
\end{figure}

Among thiobases, the series of thiouracils, i.e., 2-thiouracil (2TU), 4-thiouracil (4TU) and 2,4-dithiouracil (DTU), are particularly interesting. 
Under experimental conditions, all three compounds exist only in their oxothione or dithione tautomeric form\cite{Rostkowska1990JACS} (see Fig.~\ref{fig:structures}). 
The UV absorption spectra\cite{Parkanyi1992SC,Moustafa2005PSSRE,Khvorostov2005JPCA,Vendrell-Criado2013PPS,Pollum2014JCP} and photodeactivation mechanism\cite{Vendrell-Criado2013PPS,Taras-Goslinska2014JPPA,Mai2015JPCA,Pollum2015PCCP} of 2TU have been analysed extensively; the absorption spectrum is slightly red-shifted compared to uracil and shows signs of a low-lying dark $n\pi^*$ state.\cite{Igarashi-Yamamoto1981BBA}
The UV absorption spectra of 4TU and DTU are considerably red-shifted compared to 2TU,\cite{Pollum2015PCCP,Pollum2014TCC} but their general excited-state behaviour appears to be not much different from the one of 2TU: all three molecules show sub-picosecond intersystem crossing after excitation to the first absorption band.\cite{Pollum2015PCCP}
In contrast to the UV absorption spectra, not much has been reported on the photoelectron spectra of thiouracils.
The main interest in this respect has been laid on the anions of these compounds,\cite{Dolgounitcheva2011,Li2011} because ionizing radiation creates anions on the way to single and double strand breaks in DNA.
A single study on the tautomerism of 2TU and its methyl derivates\cite{Katritzky1990} contains a static He(I) photoelectron spectrum of neutral 2TU.

Also, up to our knowledge, no time-resolved photoelectron spectroscopy (TRPES)\cite{Stolow2004CR} has been reported for these molecules, even though this methodology has been extensively applied to study the excited-state dynamics of the canonical nucleobases\cite{Vries2014TCC,Schwell2014TCC,McFarland2014NC} (see, e.g., Ref.~\onlinecite{Mai2014TCC} for a list of these experiments) and a small number of nucleobase analogous.\cite{Golan2012NC,Poully2010PCCP}
TRPES has a number of advantages with respect to other spectroscopic techniques, e.g., it allows to study the excited-state dynamics in the gas phase, while methods like transient absorption spectroscopy usually measure solvated compounds. 
Furthermore, due to very different selection rules for photoionization compared to absorption, in TRPES it is possible to probe dark states, which is very useful to elucidate the photophysical intermediates of these compounds. 
Such knowledge can be particularly relevant for thiouracils, helping to promote the use and eventual functionalization of thiobases with phototherapeutic aims.

With this in mind, in this paper we report theoretical UV-photoelectron spectra 2TU, 4TU, and DTU from the ground state equilibrium geometries. 
We describe the character of the various ionization source orbitals and how thio-substitution influences the number, intensity, and position of the bands in the low-energy spectra. 
For 2TU, we also explore the photoelectron spectra from the neutral excited states at the Frank-Condon region as well as around the stationary points on the potential energy surface of the lowest excited singlet and triplet states, complementing our recent study\cite{Mai2015JPCA} on the relaxation and intersystem crossing mechanism after UV irradiation of this molecule. 
While we do not report time-resolved photoelectron spectra here, the spectra at the stationary points indicate that single-photon UV-TRPES alone probably cannot discern the different deactivation pathways, in particular intersystem crossing, of photoexcited 2TU and should be complemented by some other method to probe the excited-state population.

\section{Methods}

\subsection{Theory}\label{sec:theory}

An efficient way to calculate static ground or excited-state UV-photoelectron spectra is to employ Dyson norms as a measure of the ionization intensities. 
In the case of single-photon ionization, the norms of Dyson orbitals have been shown to be a good approximation for relative ionization probabilities.\cite{Hudock2007JPCA,Hudock2008C,Tao2011,Spanner2012PRA}
An excellent introduction into the description of photoionization using Dyson orbitals can be found in references~\onlinecite{Oana2007JCP,Spanner2009PRA,Spanner2012PRA}, so here we only provide a brief overview.

In the following, we describe transitions from a neutral source state $i$ to an ionic state $\alpha$ as the ionization channel $i\rightarrow\alpha$. The corresponding Dyson orbital $\left|\phi_{i,\alpha}^\mathrm{D}\right\rangle$, which can be thought of as the wavefunction of the electron to be detached, is defined as
\begin{equation}
  \label{eqn:defDysonOrb}
  \left|\phi_{i,\alpha}^\mathrm{D}\right\rangle:=\sqrt{n}\left\langle\Theta_{\mathrm{ion},\alpha}^{(n-1)}\middle|\Psi_{\mathrm{src},i}^{(n)}\right\rangle_{(n-1)},
\end{equation}
with the $n$-electron source wavefunction $\left|\Psi_{\mathrm{src},i}^{(n)}\right\rangle$, as well as the $(n-1)$-electron ionic wavefunction $\left|\Theta_{\mathrm{ion},\alpha}^{(n-1)}\right\rangle$.
The subscript $(n-1)$ of the braket indicates integration over $n-1$ electrons.
As a measure for the relative ionization yield $W_{i,\alpha}$ we use an expression analogous to the oscillator strength, where the Dyson norm is scaled with the energy difference $\Delta E_{i,\alpha}$ between the $n$-electron source state $i$ and the $(n-1)$-electron state $\alpha$:
\begin{equation}
  \label{eqn:ion_yield}
  W_{i,\alpha}\propto\Delta E_{i,\alpha} \left\langle\phi_{i,\alpha}^{\mathrm{D}}\middle|\phi_{i,\alpha}^{\mathrm{D}}\right\rangle.
\end{equation}
We do not explicitly consider the continuum wavefunction of the free electron $\left|\chi_{\alpha}^{\vec{k}}\right\rangle$ for the calculation of the intensities here.\cite{Spanner2012PRA}

In the case of wavefunctions of the configuration interaction (CI) type expanded in Slater determinants with orthonormal spin orbitals $\theta_{a,r}$ and $\psi_{b,s}$:
\begin{eqnarray*}
\left|\Theta_{\mathrm{ion},\alpha}^{(n-1)}\right\rangle&=&\sum_a c_{\mathrm{ion},\alpha,a} \left|\Theta_{\mathrm{ion},\alpha,a}^{(n-1)}\right\rangle,\\
\left|\Theta_{\mathrm{ion},\alpha,a}^{(n-1)}\right\rangle&=&\frac{1}{\sqrt{(n-1)!}}\left|\vphantom{\frac{1}{1}}\theta_{a,1}\cdots
\theta_{a,r}\cdots\theta_{a,(n-1)}\right\rangle\nonumber,\\
\left|\Psi_{\mathrm{src},i}^{(n)}\right\rangle&=&\sum_b c_{\mathrm{src},i,b} \left|\Psi_{\mathrm{src},i,b}^{(n)}\right\rangle\nonumber,\\
\left|\Psi_{\mathrm{src},i,b}^{(n)}\right\rangle&=&\frac{1}{\sqrt{n!}}\left|\vphantom{\frac{1}{1}}\psi_{b,1}\cdots
\psi_{b,s}\cdots\psi_{b,n}\right\rangle,\nonumber
\end{eqnarray*}
the total Dyson orbital is the sum of the overlaps of all Slater determinant pairs (indices $a$, $b$) weighted with the product of their CI coefficients
\begin{equation}
  \left|\phi_{i,\alpha}^{\mathrm{D}}\right\rangle_{\mathrm{CI}}=\sum_a \sum_b c_{\mathrm{ion},\alpha,a}^* c_{\mathrm{src},i,b} \left|\phi_{i,\alpha}^{\mathrm{D}}\right\rangle_{a,b}.
\end{equation}
To this end, the contribution of one Slater determinant pair  $\left\langle\Theta_{\mathrm{ion},\alpha,a}^{(n-1)}\right|$ and $\left|\Psi_{\mathrm{src},i,b}^{(n)}\right\rangle$ can be rewritten in terms of the molecular spin-orbitals of the source wavefunction, $\left|\psi_{b,s}\right\rangle$, and the annihilation operator, $\hat{\mathbf{a}}_s$, for the $s$-th occupied source spin-space orbital, $\left|\psi_{b,s}\right\rangle$:
\begin{eqnarray}
  \left|\phi_{\alpha}^{\mathrm{D}}\right\rangle_{a,b}&=&\sqrt{n}\left\langle\Theta_{\mathrm{ion},\alpha,a}^{(n-1)}\middle|\Psi_{\mathrm{src},i,b}^{(n)}\right\rangle_{(n-1)}\\
  &=&\sqrt{n}\sum_s^{occ_{\mathrm{src}}}\left\langle\Theta_{\mathrm{ion},\alpha,a}^{(n-1)}\middle|\hat{\mathbf{a}}_s\middle|\Psi_{\mathrm{src},i,b}^{(n)}\right\rangle_{(n-1)}\left|\psi_{b,s}\right\rangle\nonumber
\end{eqnarray}
The overlap $\left\langle\Theta_{\mathrm{ion},\alpha,a}^{(n-1)}\middle|\hat{\mathbf{a}}_s\middle|\Psi_{\mathrm{src},i,b}^{(n)}\right\rangle_{(n-1)}$ can be then obtained as a matrix determinant over appropriate molecular orbital overlap integrals.

%%=================================================================================================
%%=================================================================================================

\subsection{Computational Details}\label{sec:compdet}

The source and ionic wavefunctions have been calculated with the quantum-chemistry package \textsc{Columbus 7.0}~\cite{Lischka2011WCMS,COLUMBUS-program2012} using atomic integrals from \textsc{Molcas 8.0}\cite{Aquilante2015JCC}. The cc-pVDZ basis set\cite{DunningJr1989,Woon1993} was used in all calculations. Neutral and ionic wavefunctions were computed on the multireference configuration interaction with single excitations (MRCIS) level of theory using orbitals taken from preceding complete active space self-consistent field (CASSCF) calculations. The active space comprised 14 electrons (13 for doublets) in 10 orbitals.
To describe ionization from the singlet states, state-averaging (SA) with equal weights has been performed using 4 states for the neutral singlet and 10 states for the ionized doublet wavefunctions, denoted as SA(4S) and SA(10D), respectively. To describe ionization from the triplet states, the neutral and ionized states employed a SA(4S+4T) and SA(10D+10Q) averaging protocol, respectively.

In the MRCIS calculation, the active orbitals of the preceding CASSCF have been used as reference space, restricting the references to configurations with at most two electrons in the three antibonding $\pi^*$ orbitals. In the MRCIS procedure, 12~orbitals for 2TU and 4TU and 16~orbitals for DTU were kept frozen. A total of 12 doublet and quartet (if applicable) states were computed on the MRCIS level of theory. This level of theory cannot accurately describe singlet and triplet excitation energies due to a deficiency of dynamic correlation (see below). However, having all important ionization source orbitals in the reference space ensures an adequate description of the lower ionized states.

Using the so-obtained wavefunctions, Dyson norms have been calculated to estimate photoionization probabilities. 
To reduce the computational cost, the CI vectors were truncated by removing the configuration state functions with the smallest absolute coefficients until the remaining wavefunction has a norm of 0.97.\cite{Plasser2015}
A careful inspection showed that this procedure strongly reduces the computational costs of the Dyson norms without significant deterioration of the results.

Finally, photoelectron spectra have been calculated from the $\mathrm{S}_0$ minimum and, in case of 2TU also from the $\mathrm{S}_1$ and $\mathrm{T}_1$ minima, optimized at the MRCIS level of theory specified above.
In order to simulate vibrational broadening, for each compound 200 geometries were sampled randomly from a Wigner distribution based on frequencies obtained at the MRCIS level of theory. For each geometry, the Dyson norms for all applicable pairs of states were computed using the molecular orbital coefficients and CI vectors from the corresponding quantum-chemistry calculations. Spectra were calculated as a sum over Gaussians centered at the excitation energies and height proportional to $W$ in equation~\eqref{eqn:ion_yield}. A full width at half maximum of $0.2\,\mathrm{eV}$ was chosen for the purpose of obtaining a smoothed spectrum from the calculated number of geometries. 
The photoelectron spectrum of the ground state of 2TU was normalized to a maximum signal of one, and the same normalization factor was then used for all other spectra (Note that all spectra were calculated with the same number of geometries and the same full width at half maximum).
The generation of photoelectron spectra has been performed using tools from the development version of the \textsc{Sharc} molecular dynamics suite.\cite{Richter2011JCTC,Mai2014SHARC,Mai2015IJQC}

%%=================================================================================================
%%=================================================================================================

\section{Results and Discussion}\label{sec:results}

\subsection{Photoelectron spectra of 2TU, 4TU and DTU}\label{ssec:CompareIsomeres}

\begin{figure}
  \begin{center}
    \includegraphics[scale=1]{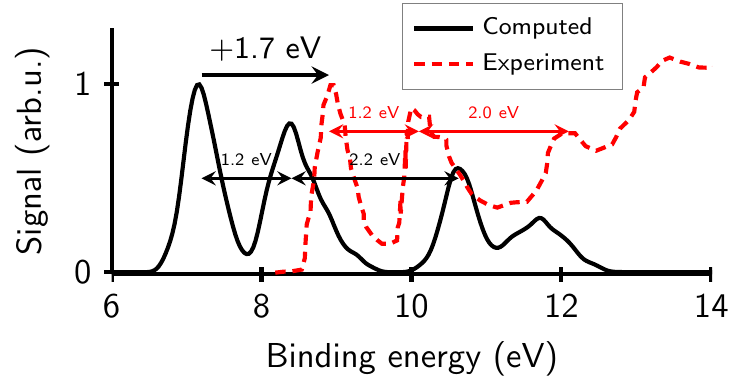}
  \end{center}
  \caption{Computed $\mathrm{S}_0$-photoelectron spectrum for 2TU compared to the experimental spectrum from Katritzky et al.\cite{Katritzky1990} Arrows indicate the band spacing and the shift of $+1.7$~eV needed to overlap the simulated spectrum with the experimental one.}
  \label{fig:2TU_pes_experiment}
\end{figure}

The photoelectron spectrum of 2TU computed from the electronic ground state is compared to the experimental one of Katritzky et al.\cite{Katritzky1990} in Figure~\ref{fig:2TU_pes_experiment}. As can be seen, there is a shift of $+1.7\,\mathrm{eV}$ towards lower binding energies (as used here, the vertical energy difference between the neutral source state and the ionic state under consideration) with respect to the experimental values. This shift originates from the lack of size extensivity\cite{Bartlett1981ARPC} of MRCIS and the difference in the number of correlated electrons in neutral and ionized wavefunctions. However, the peak separations and relative peak heights in the simulated spectrum agree nicely with the experimental data, particularly at low energies, showing that the method predicts the energetic spacing of the ionic states and the relative intensities correctly.
For higher binding energies the experimental spectrum shows a very broad band starting at $\approx11\,\mathrm{eV}$, which could not be reproduced by our calculations. We assume that this band originates either from ionization to higher-lying excited doublet states or from ionization from $\sigma$ orbitals, both effects not well described by our simulation.
An extension of the simulated energy range to higher doublet states with the present level of theory is not expected to yield better results because already in the present calculations the highest states ($\mathrm{D}_{9-11}$) are strongly mixed and have large contributions of double excitations within the reference space. Since the description of higher-lying states would require a much more accurate electronic structure method, which is out of the scope of this paper, the analysis following in the next sections focuses on the lower part of the spectrum.

\begin{figure}
  \begin{center}
    \includegraphics[scale=1]{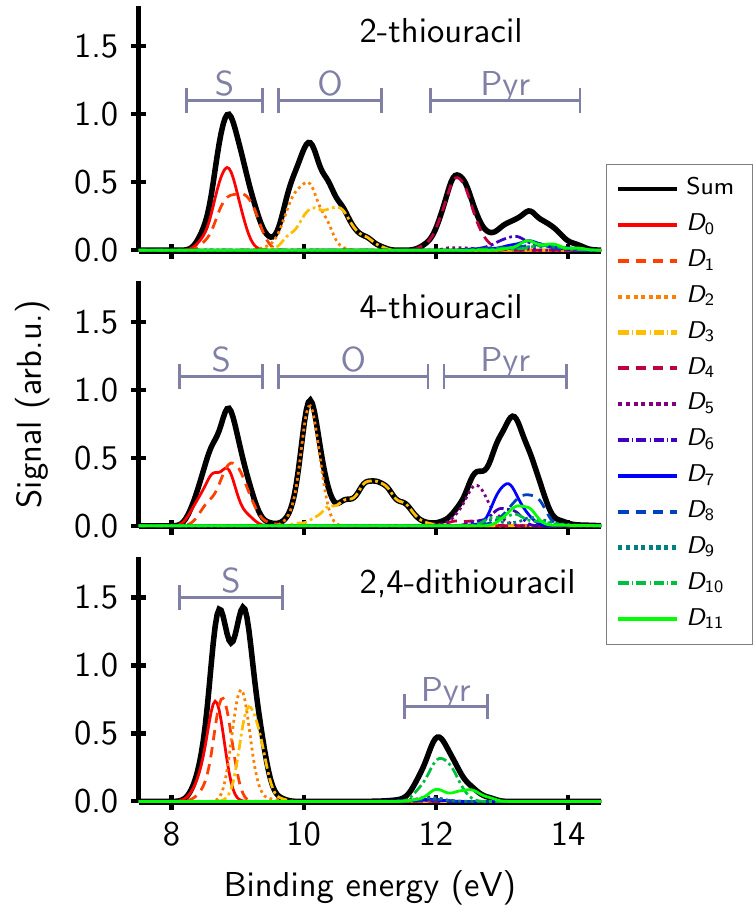}
  \end{center}
  \caption{Computed $\mathrm{S}_0$-photoelectron spectra for 2TU, 4TU and DTU with individual contributions of the first 12 ionization channels ($\mathrm{D}_{0}$-$\mathrm{D}_{11}$). ``S'', ``O'' and ``Pyr'' labels indicate predominant ionization from orbitals on sulphur, oxygen and the pyrimidine ring respectively. All spectra have been shifted by $+1.7\,\mathrm{eV}$ (off-setting the shift shown in Figure~\ref{fig:2TU_pes_experiment}).}
  \label{fig:S0_spectra_all}
\end{figure}
\begin{figure*}
  \includegraphics[scale=1]{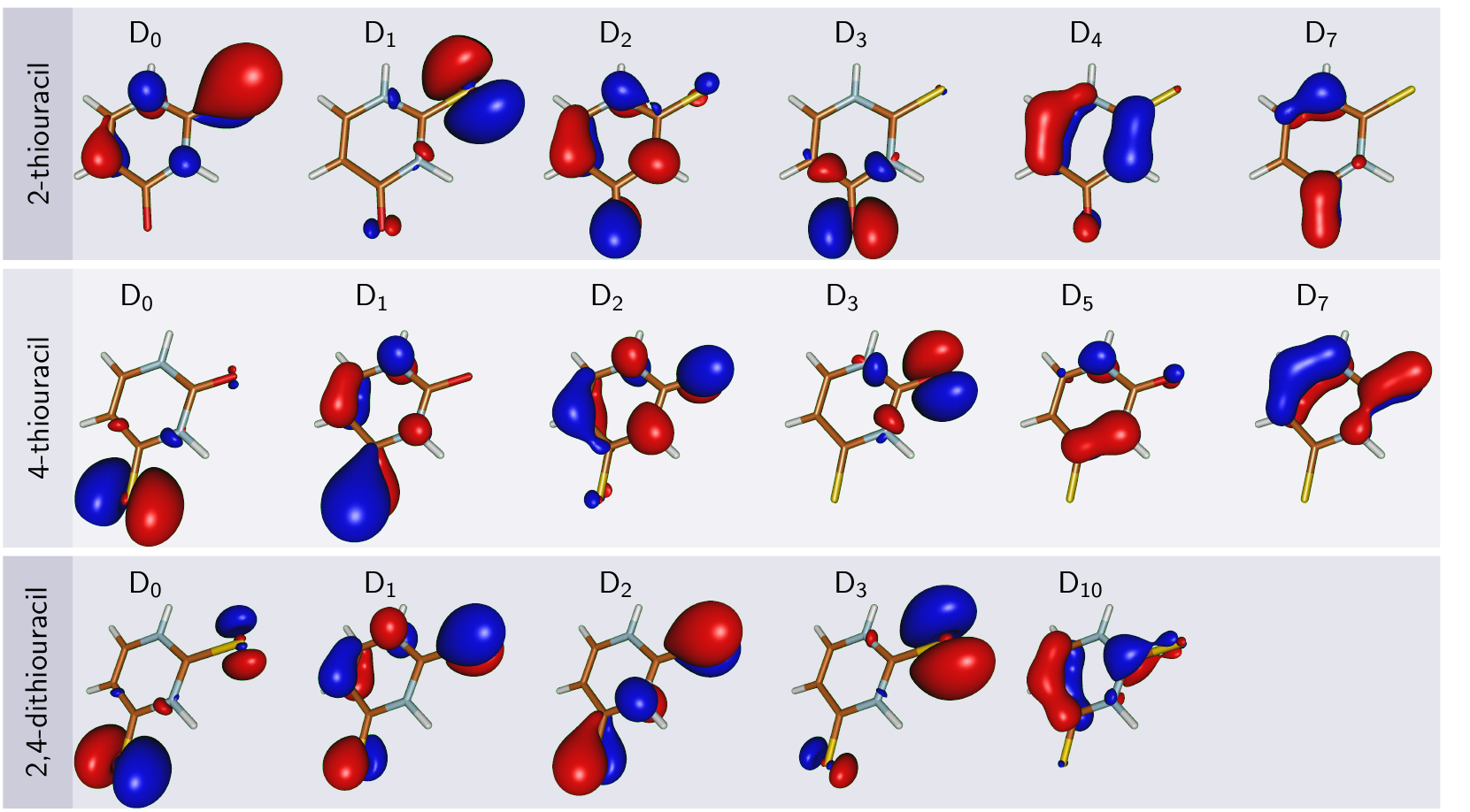}
  \caption{Dyson orbitals for the relevant ionizations channels from the $\mathrm{S}_0$ state of the three thiouracils. The orientation of the molecules is the same as in Figure~\ref{fig:structures}.}
  \label{fig:S0_dysorbs_all}
\end{figure*}

The simulated photoelectron spectra of the neutral ground states of all three compounds are presented in Figure~\ref{fig:S0_spectra_all}, together with the contributing signals of the ionization channels ($\mathrm{D}_{0}$-$\mathrm{D}_{11}$).
While 2TU and 4TU both show three groups of signals, the spectrum of DTU is characterized by only two bands. The band lowest in energy ($\approx8\,\mathrm{eV}-10\,\mathrm{eV}$) is common to all three compounds but it is more intense in DTU than in the other two molecules due to the different composition of the bands. Only the $\mathrm{D}_{0}$ and $\mathrm{D}_{1}$ ionization channels contribute to the first band of 2TU and 4TU, whereas $\mathrm{D}_{0}$-$\mathrm{D}_{3}$ compose the first band of DTU. The next band higher in energy, centered around $10\,\mathrm{eV}$ is only present for 2TU and 4TU while there is no visible intensity for DTU. The shape of this band, which is different for 2TU compared to 4TU, is due to the different intensities and energies of the contributing $\mathrm{D}_{2}$ and $\mathrm{D}_{3}$ channels. The third band beyond $12\,\mathrm{eV}$ is different for each of the molecules.

An analysis of the ionization channels in terms of their corresponding Dyson orbitals, which describe the character of the departing electron's wavefunction, is very instructive to understand the differences between the spectra. Figure~\ref{fig:S0_dysorbs_all} shows all Dyson orbitals with norm $>0.1$ for the relevant ionization channels present in Figure~\ref{fig:S0_spectra_all}.
As can be seen, the $n$ and $\pi$ orbitals of the sulphur atom are the dominant contributions to the Dyson orbitals of the $\mathrm{S}_0\rightarrow\mathrm{D}_{0,1}$ ionizations in 2TU and 4TU, as well as of the $\mathrm{S}_0\rightarrow\mathrm{D}_{0-3}$ ionizations in DTU. Accordingly, the first band of the three spectra is due to ionization from sulphur.
The Dyson orbitals of the $\mathrm{S}_0\rightarrow\mathrm{D}_{2,3}$ transitions in 2TU and 4TU correspond mostly to ionization from the $n$ and $\pi$ orbitals of oxygen. The different extent of mixing of the oxygen $\pi$ orbital with pyrimidine $\pi$ orbitals explains the differently shaped bands in 2TU and 4TU.
Since there is no oxygen in DTU, it is apparent that this second band around $10\,\mathrm{eV}$ binding energy is missing in this compound.
Finally, the signals in the high-energy range ($>12\,\mathrm{eV}$) contain the contributions of ionization from the $\pi$ orbitals of the pyrimidine ring. The binding energy for electrons involved in a bond is naturally higher than for the lone pairs. The density of doublet states is very high in this energy region and many states show no significant intensity. For DTU only the $\mathrm{D}_{10}$ yields substantial ionization probability. Hence, we assume that this band is only partially described by our calculations.

The simulated spectra show that the three compounds, 2TU, 4TU and DTU could certainly be discriminated on the basis of their photoelectron spectra. DTU is clearly marked by the intensity of the sulphur-ionization band and the absence of an oxygen-ionization band. In contrast, 2TU and 4TU show distinctly different shapes of the band arising from ionization of oxygen-localized orbitals.

%%=================================================================================================

\subsection{Excited-state photoelectron spectra of 2TU }\label{ssec:2TU-excst}

\begin{figure}
  \centering
  \includegraphics[scale=1]{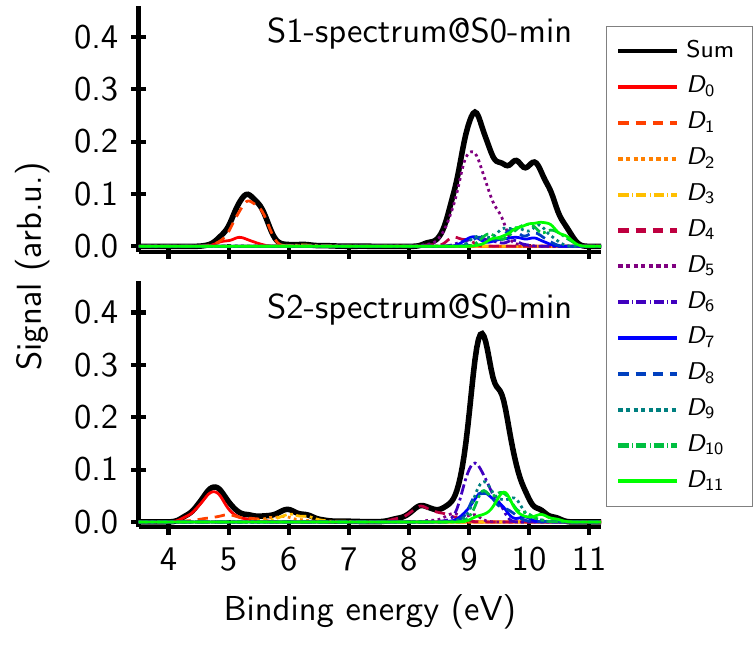}
  \caption{Calculated $\mathrm{S}_1$ ($n\pi^*$)- and $\mathrm{S}_2$ ($\pi\pi^*$)-photoelectron spectra of 2TU including the contributions of all computed ionization channels. The spectra have been shifted by $(1.7+0.5)\,\mathrm{eV}$ and $(1.7+0.77)\,\mathrm{eV}$ for for $\mathrm{S}_1$ and $\mathrm{S}_2$, respectively.}
  \label{fig:S1S2_spectra_2TU}
\end{figure}

\begin{figure}
        \centering
  \includegraphics[scale=1]{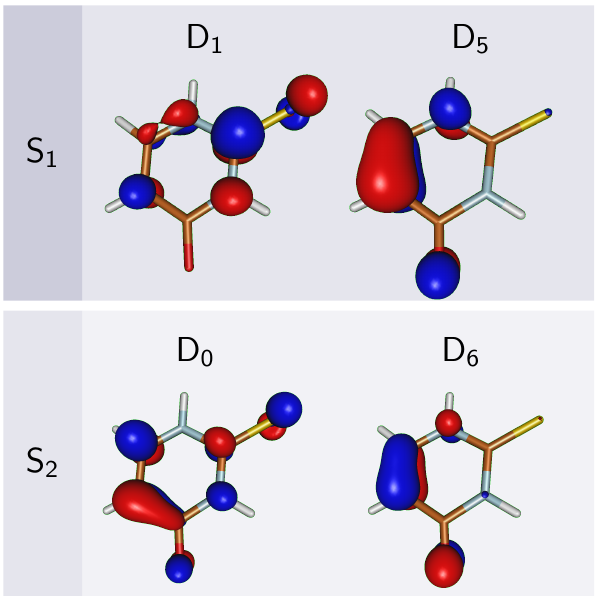}
  \caption{Dyson orbitals for the important low-energy ionization channels from the $\mathrm{S}_1$ and $\mathrm{S}_2$ states of 2TU.}
  \label{fig:S1S2_dysorbs_2TU}
\end{figure}

\begin{table}
  \centering
  \caption{Calculated vertical excitation energies (eV) and oscillator strengths (in parentheses) of $\mathrm{S}_1$ and $\mathrm{S}_2$ and relative energies of $\mathrm{S}_1$/$\mathrm{T}_1$ minima of 2TU at the MRCIS and CASPT2 levels of theory, the latter taken from Ref.~\onlinecite{Mai2015JPCA}.}
  \begin{tabular}{llcc}
    State  &Geometry       &This work &CASPT2(16,12)\cite{Mai2015JPCA}  \\
    \hline
    $\mathrm{S}_1$ ($n_S\pi^*$)   &$S_0$ min                & $4.27$ ($0.00$) & $3.77$ ($0.00$) \\
    $\mathrm{S}_2$ ($\pi_S\pi^*$) &$S_0$ min                & $5.02$ ($0.29$) & $4.25$ ($0.35$)  \\
    \hline
    $\mathrm{S}_1$ ($n_S\pi^*$)   &$S_1$ min                & $3.79$          & $3.45$ \\
    $\mathrm{T}_1$ ($\pi_S\pi^*$) &$T_1$ ``boat'' min       & $3.36$          & $3.35$ \\
    $\mathrm{T}_1$ ($\pi_S\pi^*$) &$T_1$ ``pyr.'' min       & $3.43$          & $3.21$
  \end{tabular}
  \label{tab:2TU_singlet_states}
\end{table}

Using the same set of geometries as for the 2TU $\mathrm{S}_0$ photoelectron spectrum, the excited-state (source states $\mathrm{S}_1$ and $\mathrm{S}_2$) photoelectron spectra have been calculated (see Figure~\ref{fig:S1S2_spectra_2TU}). 
The character of the neutral excited states at the equilibrium geometry is listed in Table~\ref{tab:2TU_singlet_states}, together with vertical excitations energies and oscillator strengths, calculated at the MRCIS level of theory and compared to our recently published CASPT2 results.\cite{Mai2015JPCA}
Irrespective of the level of theory, the order of the states is the same and the $\mathrm{S}_1$ state is an $n\pi^*$ excitation while the $\mathrm{S}_2$ is a $\pi\pi^*$ state. In both states, an electron is excited from an orbital located at the sulphur atom and promoted to a $\pi^*$ orbital on the pyrimidine ring.
Quantitatively, one can see that the excitation energies are described too high at the MRCIS level of theory, more so for the $\pi\pi^*$ state. 
As this would lead to systematically too low binding energies in the simulation of photoelectron spectra with excited states as ionization source, an additional shift of $+0.5\,\mathrm{eV}$ (towards higher binding energies) for the $\mathrm{S}_1$ spectrum and $+0.77\,\mathrm{eV}$ for the $\mathrm{S}_2$ one was introduced.

Ionization of an electron which has been excited into an antibonding orbital (e.g., $\pi^*$) requires lower energy than to eject electrons in bonding orbitals. Consequently, the Dyson orbitals (Figure~\ref{fig:S1S2_dysorbs_2TU}) for the first ionizations ($\mathrm{S}_1\rightarrow\mathrm{D}_{1}$ and $\mathrm{S}_2\rightarrow\mathrm{D}_{0}$) are very similar to the $\pi^*$ orbitals occupied in the $\mathrm{S}_1$ and $\mathrm{S}_2$ states.
The second band arises mostly from ionization of the $\pi$ orbital located at the $C_5=C_6$ bond.

The first band of each spectrum (Figure~\ref{fig:S1S2_spectra_2TU}) is in the region of $4\,\mathrm{eV}-6\,\mathrm{eV}$ and thus more than $2\,\mathrm{eV}$ separated from the onset of the $\mathrm{S}_0$ spectrum and from the next band in the excited-state spectra. Judging from the simulated spectra it should be possible to obtain a clear signal from the lowest ionization channels of the excited singlet states without interference of the ground state signal using probe laser energies of $6\,\mathrm{eV}-8\,\mathrm{eV}$. This allows for experimental studies using photoelectron spectroscopy to selectively track excited states. 
Furthermore, the spectra in Figure~\ref{fig:S1S2_spectra_2TU} show that $\mathrm{S}_1$ and $\mathrm{S}_2$ are distinguishable in the Frank-Condon region, but it is expected that during the dynamics after excitation their spectra will broaden and shift independently and hence might not be separable when moving away from the Franck-Condon point.

%%=================================================================================================

\begin{figure}
  \begin{center}
    \includegraphics[scale=1]{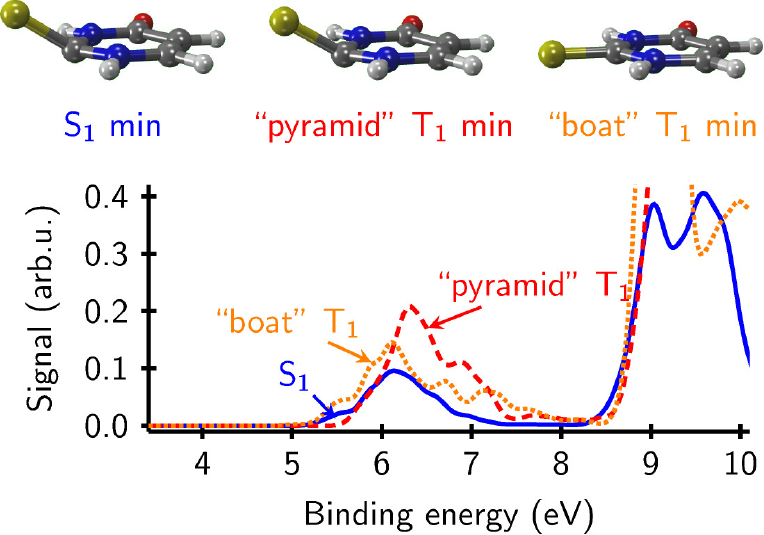}
  \end{center}
  \caption{Calculated photoelectron spectra of 2TU from the $\mathrm{S}_1$ at the $\mathrm{S}_1$ minimum and from the $\mathrm{T}_1$ at the two $\mathrm{T}_1$ minima (boat and pyramidalized conformations\cite{Mai2015JPCA}). On the top, the minimum geometries are presented. The spectra have been shifted by $+1.7\,\mathrm{eV}$.}
  \label{fig:S1minT1min_compare_2TU}
\end{figure}

An important application of photoelectron spectroscopy in combination with pump-probe experiments is the ability to track different states and relaxation pathways in the excited-state dynamics. A recent exploration of the possible reaction pathways using static quantum-chemical calculations\cite{Mai2015JPCA} suggests that 2TU will, after electronic excitation to the $\mathrm{S}_2$, quickly cross to the $\mathrm{S}_1$ where it is trapped in a minimum with low chance to reach the crossing seam to the ground state. Instead, it is very likely that the system undergoes intersystem crossing into the energetically close $\mathrm{T}_{2}$ state and from there internally converts into one of two $\mathrm{T}_1$ minima (named ``boat'' and ``pyramidalized'' in Ref.~\onlinecite{Mai2015JPCA}).
An interesting question is to investigate whether it will be possible to gain insight into this intersystem crossing process of 2TU by TRPES. We therefore optimized the $\mathrm{S}_1$ and both $\mathrm{T}_1$ minima (geometries shown in Figure~\ref{fig:S1minT1min_compare_2TU}) and generated photoelectron spectra from these minima. 

Table~\ref{tab:2TU_singlet_states} collects the relative energies of these minima with respect to the equilibrium geometry at the MRCIS and MS-CASPT2 levels of theory. Despite the energetic order of the two $\mathrm{T}_1$ minima being different with both methods, they are still very close in energy and no additional shift was applied. 
Figure~\ref{fig:S1minT1min_compare_2TU} displays the three excited-state photoelectron spectra of 2TU. For each minimum, we obtain a weak band in the 5--8~eV energy range and an intense band above 8.5~eV, where the latter band is much more intense for the triplet state due to ionization to quartet states.
The low-energy bands are quite similar and strongly overlapping and the high-energy band will likely be covered by ionization from the ground state.
Hence, based on these spectra, we suggest that the $\mathrm{S}_1$ and $\mathrm{T}_1$ states cannot be discriminated by single-photon ionization using a probe laser in the suggested energy range of $6\,\mathrm{eV}-8\,\mathrm{eV}$. Depending on the actual probe laser energy used, only part of the bands will contribute to the signal and hence a low intensity can be expected. Most importantly, the bands of the three spectra overlap strongly, and as can be seen, the small error in the energies of the neutral states due to MRCIS is not relevant. Based on these results, we predict that single-photon TRPES alone will not allow to track the population transfer from $\mathrm{S}_1$ to $\mathrm{T}_1$ in 2TU.

\section{Conclusions}

The UV-photoelectron spectra of three thio-analogues of uracil, 2-thiouracil (2TU), 4-thiouracil (4TU) and 2,4-dithiouracil (DTU) have been simulated using Dyson norms as a measure for the relative ionization probabilities of the different ionization channels. 
In all three molecules, the respective first band of the photoelectron spectra is mainly due to ionization from the $n$ and $\pi$ orbitals on sulphur.
For 2TU and 4TU, the second band corresponds to ionization from the $n$ and $\pi$ orbitals on oxygen, while for all three compounds the highest calculated band is related to the pyrimidine $\pi$ system.
These two (DTU) or three (2TU, 4TU) bands are well separated. 

In the monosubstitued thiouracils (2TU and 4TU) the position of the substitution has some effect on the band shape. With reasonably good resolution in the energy domain, we predict that the two molecules could be discerned mainly by the shape of the bands that originate from ionization from the oxygen orbitals at around 10~eV binding energy.
The photoelectron spectrum of DTU is easily distinguishable from the other compounds since it lacks the band from the oxygen-orbital ionization completely and in return it shows an intense double peak from the ionization out of orbitals located on the sulphur atoms.

An inspection of the excited-state photoelectron spectra from the $\mathrm{S}_1$ and $\mathrm{S}_2$ states of 2TU shows that the excited states can be detected without being disturbed by the ground state signal.
Furthermore, the photoelectron spectra of 2TU simulated at the relevant $\mathrm{S}_1$ and $\mathrm{T}_1$ minima show strong overlap and very similar intensity.
Hence, experimental pump-probe UV-TRPES using single-photon ionization alone might not be able to track the populations of the different excited singlet and triplet states and therefore intersystem crossing could be opaque to this methodology.
However, other experimental techniques like multi-photon ionization might still be able to detect a signature from intersystem crossing.

%%=================================================================================================

\section*{Acknowledgments}

This work was financed by the Austrian Science Fund (FWF) project P25827.
Part of the calculations were performed on the Vienna Scientific Cluster 3 (VSC3).
The authors want to thank Esther Heid for preliminary calculations, Susanne Ullrich and Carlos Crespo-Hern\'andez for bringing up questions related to photoelectron spectra on thiouracils and the COST Action CM1204 (XLIC) for stimulating discussions.
We also want to thank Serguei Patchkovskii and Michael Spanner for discussions and for providing their code for Dyson orbital calculations which helped our development very much.

%%=================================================================================================
% \bibliography{matruc_bibfile}


\end{document}